\documentclass[aps,12pt,
		prd,
		reprint,
		onecolumn,
		superscriptaddress,
		shortbibliography,
		nofootinbib,
		floatfix,
		notitlepage
		]{revtex4-1}

\usepackage{amsmath,amssymb,amsfonts}

\usepackage{natbib}
\usepackage[T1]{fontenc}
\usepackage[latin1]{inputenc}

\usepackage[dvipsnames]{xcolor}

\usepackage{hyperref}
\hypersetup{colorlinks=true,citecolor=Cyan,urlcolor=Red}

\usepackage{mathrsfs}
\usepackage{bbm}
\usepackage{slashed}
\DeclareSymbolFont{rsfs}{U}{rsfs}{m}{n}
\DeclareSymbolFontAlphabet{\mathrsfs}{rsfs}
\usepackage[mathscr]{eucal}	
\usepackage[normalem]{ulem}
\usepackage{verbatim}
\usepackage{bm}

\usepackage{graphicx}

\newcommand{\be}{\begin{equation}}
\newcommand{\ee}{\end{equation}}
\newcommand{\bi}{\begin{itemize}}
\newcommand{\ei}{\end{itemize}}
\newcommand{\bea}{\begin{eqnarray}}
\newcommand{\eea}{\end{eqnarray}}
\newcommand{\ud}{\mathrm{d}}		

\newcommand{\LCm}{{\scriptscriptstyle -}} 
\newcommand{\LCp}{{\scriptscriptstyle +}}

\newcommand{\LCperp}{{\scriptscriptstyle \perp}}

\begin{document}
%


\title{Exact results for scattering on ultra-short plane wave backgrounds}

\author{A.~Ilderton}
\email{anton.ilderton@plymouth.ac.uk}
\affiliation{Centre for Mathematical Sciences, University of Plymouth, Plymouth, PL4 8AA, UK}

\begin{abstract}	
We give exact results for the emission spectra of both nonlinear Breit-Wheeler pair production and nonlinear Compton scattering in ultra-intense, ultra-short duration plane wave backgrounds, modelled as delta-function pulses. This includes closed form expressions for total scattering probabilities. We show explicitly that these probabilities do not exhibit the power-law scaling with intensity associated with the conjectured breakdown of (Furry picture) perturbation theory, instead scaling logarithmically in the high-intensity limit. 
\end{abstract}
\maketitle

\section{Introduction}
%
The coupling of matter to a background field can be arbitrarily strong, requiring non-perturbative methods in the calculation of scattering amplitudes on the background. Highly symmetric backgrounds such as electromagnetic plane waves~\cite{volkov35,Bagrov:1990xp,Heinzl:2017zsr}, widely used as a first model of intense laser fields~\cite{RitusReview,DiPiazza:2011tq,King:2015tba,Seipt:2017ckc}, may be treated exactly for arbitrarily high field strength, allowing a great deal of analytic progress to be made in the calculation of scattering amplitudes.

However, the evaluation of observables still requires intense numerical integration. While various results have been found which allow more efficient calculation~\cite{Dinu:2013hsd,Seipt:2016rtk,Dinu:2018efz}, it would be desirable to have exact results when confronting questions of a fundamental nature, such as the limits of perturbation theory in the high intensity regime; it has been conjectured that (Furry picture~\cite{Furry51}) perturbation theory in strong fields breaks down for sufficiently high field strengths, due to a power-law scaling of higher loop processes~\cite{Ritus1,Morozov1,Narozhnyi:1979at,Narozhnyi:1980dc,Morozov2}. If true, this would invalidate current perturbative or semi-perturabtive approaches to QED in strong backgrounds~\cite{Fedotov:2016afw,SLACKonf}.

Here we have two objectives. The first is to provide some exact results for scattering on ultra-short plane wave pulses. This includes closed-form expressions for total probabilities. Our second objective is to investigate the high-intensity scaling of these results in light of the conjectured high-intensity breakdown. The model on which this conjecture is based, and the starting point for many investigations of laser-matter interactions, assumes the laser fields may be treated as constant and homogeneous~\cite{RitusReview}. We consider here the opposite limit in which the field is of infinitely short duration, being modelled as a sequence of delta function pulses. This will allow us to provide exact results: see also~\cite{Fedotov:2013uja} for the use of delta pulses, in combination with constant fields, to Schwinger pair production, where exact results are also obtained. 

This paper is organised as follows. In Sect.~\ref{SECT:ALLM} we review some necessary structures of QED scattering in plane wave backgrounds. In Sect.~\ref{SECT:DC} we consider stimulated pair production in a delta-function pulse.  Both the differential emission spectra and the total pair creation probability are obtained in closed form, with all integrals performed. We show that the high-intensity scaling of the probability is logarithmic, not power law. We then consider the case of two delta function pulses modelling an oscillating field. In Sect.~\ref{SECT:NLC} we turn to nonlinear Compton scattering, and find that the leading order, high intensity, behaviour in an oscillating field is doubly logarithmic. We conclude in Sect.~\ref{SECT:CONCS}.

\subsection{General structures}\label{SECT:ALLM}
Consider the scattering of some collection of particles incident on a background plane wave. The wave is described by the potential $eA_\mu = a_\mu(n.x)$ where $n_\mu$ is null, i.e.~$n^2=0$. We work in lightfront coordinates, $x^\LCp := x^0 \pm x^3$ and $x^\LCperp =\{x^1,x^2\}$. We can always choose $n.x=x^\LCp$. We can also choose the potential to have only nonzero ``perpendicular''  components $a_\LCperp$, and to obey $a_\LCperp(-\infty)=0$~\cite{Dinu:2012tj}; the derivative of $a_\LCperp$ then gives the electric field of the wave, while $a_\LCperp$ itself describes the work done on a particle entering the wave from $n.x = -\infty$~\cite{Brodsky:1997de,Heinzl:2000ht,Bakker:2013cea}.

$S$-matrix elements in plane wave backgrounds take the form
\be
	S_{fi} = \frac{1}{2} (2\pi)^3 \delta^3_{\LCm,\LCperp}(p_f - p_i)\, \mathcal{M} \;,
\ee
in which $p_i$ and $p_f$ represent sums of initial and final particle momenta, the delta-function conserves the three momentum components $p_\LCm:=(p_0-p_3)/2$ and $p_\LCperp:=\{p_1, p_2\}$, and the nontrivial part $\mathcal{M}$ has the structure, for the cases of interest here,
\be\label{MFI-0}
	\mathcal{M} = \int\!\ud\phi\, e^{i\Phi} \, \text{Spin}(a) \;,
\ee
in which $\phi\equiv n.x$, $\Phi$ is a real function of $\phi$ with nonvanishing $\phi$-derivative (denoted $\Phi'$), and ``Spin'' is a combination of spin and polarisation structures which depends on $\phi$ through the background $a_\mu(\phi)$.  The explicit forms of $\Phi$ and Spin are calculated in the Furry picture, which treats the background exactly, using the well-known Volkov solutions in plane wave backgrounds. As such calculations are commonplace in the literature, we do not dwell on the details; see e.g.~\cite{Seipt:2017ckc} for an introduction. The only subtlety is that we use the modified LSZ prescription detailed in~\cite{Kibble:1965zza,Dinu:2012tj} in order to account for unipolar fields for which the potential may not vanish asymptotically

We consider processes which cannot occur in the absence of the field, which means that $\mathcal{M}$ must vanish as the field is switched off. While this will be formally true of (\ref{MFI-0}), it is convenient to make it explicit by, essentially, symbolically subtracting the $a_\mu\to 0$ result~\cite{Boca:2009zz} (in a manner consistent with gauge-invariance~\cite{Ilderton:2010wr}). This leads to the following ``regulated'' expression for $\mathcal{M}$ in any pulsed plane wave~\cite{Dinu:2012tj};
\be\label{DC-REG-1}
	\mathcal{M} = -\int\!\ud\phi \, e^{i\Phi} \frac{\ud}{\ud\phi} \bigg[\frac{\text{Spin}(a)}{i\Phi'}\bigg] \;.
\ee
Integrating by parts, the boundary term will in general contribute.

Several kinds of integral arise in the calculation of scattering probabilities in plane waves. Integrals over transverse final state momenta are always Gaussian~\cite{Dinu:2013hsd}, and integrals over longitudinal momenta (of form $n.p$) can often be performed to give special functions~\cite{Dinu:2013hsd,Dinu:2013gaa}. However the $\phi$-integral in $\mathcal{M}$ cannot be performed analytically in general. There are two special cases, constant and monochromatic fields, for which symmetry allows the $\phi$-integral to be performed. However, the resulting change in form then prohibits the final state momentum integrals from being performed. In contrast to this, we will in the examples below be able to perform the $\phi$-integrals, and more, exactly.
%
\section{Pair production in delta-function pulses} \label{SECT:DC}
%
\begin{figure}[t]
\includegraphics[width=0.3\textwidth]{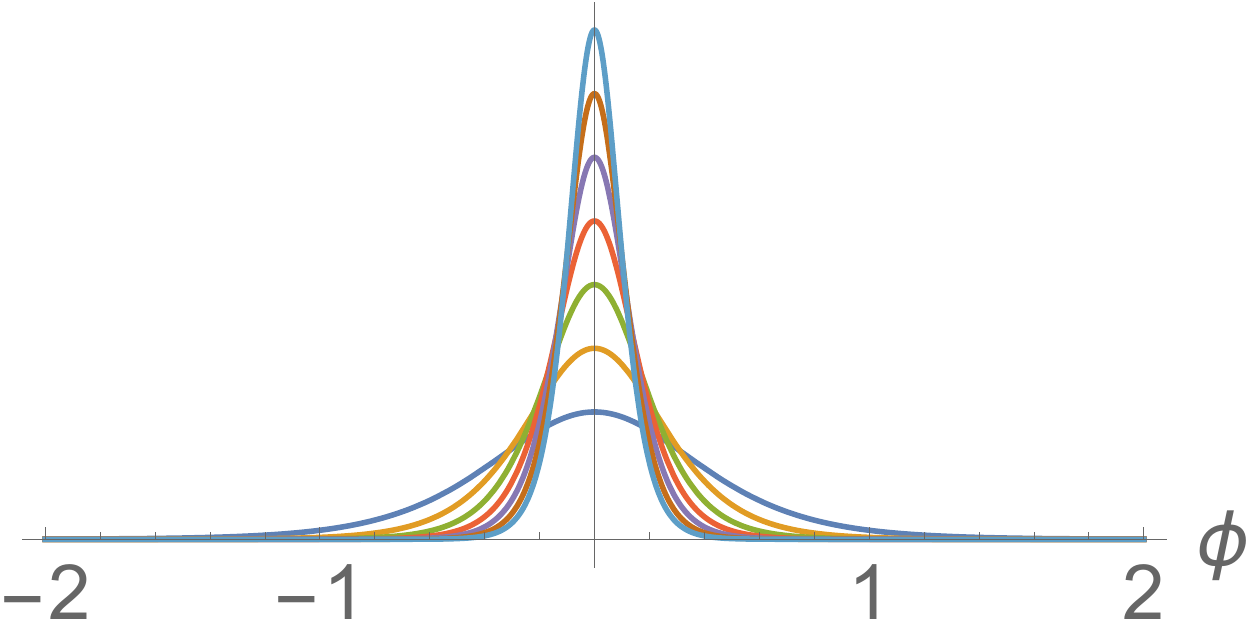} \qquad \qquad \qquad \qquad
\includegraphics[width=0.3\textwidth]{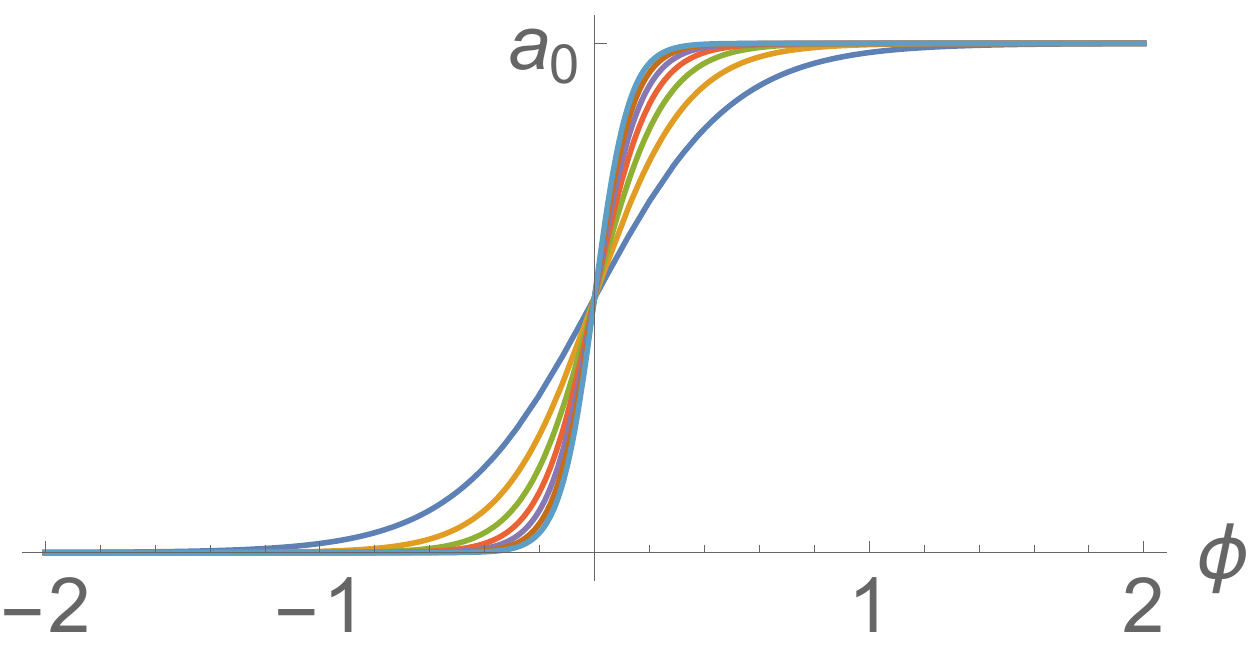}
\caption{\label{FIG:MED-DC} The electric field (left) and potential (right) for a unipolar field. In the considered limit, the electric field becomes a delta function at the origin, while the potential becomes a Heaviside step function.}
\end{figure}
We consider the production of an electron-positron pair, momenta $p_\mu$ and $q_\mu$ respectively, from a photon of momentum $l_\mu$ incident on a plane wave pulse. We take the electric field of the pulse to be a single peak as illustrated in~Fig.~\ref{FIG:MED-DC}. Such fields, called unipolar~\cite{Kozlov} or impulsive, transfer net energy-momentum to a classical particle traversing them, via a nonzero Fourier zero mode of the electric field~\cite{Dinu:2012tj}. (See~\cite{Aichelburg1971,Penrose:1972xrn,Dray,Klimcik:1988az,Ferrari1988,Adamo:2017nia} for various applications in gravity.) The specific case of interest here is a linearly polarised electric field with $E_2=a_2=0$ and
\be\begin{split}\label{LIMITS0}
	E_1(\phi) &= \frac{m\omega a_0}{2} \text{sech}^2 (\omega \phi)  \;, \qquad 
	a_1(\phi) = m a_0 \Big(1 + \tanh(\omega \phi)\Big) \;.
\end{split}
\ee
In the limit $\omega\to\infty$, see also Fig.~\ref{FIG:MED-DC}, $E_1$ becomes a delta function and the potential becomes a step function,  
\be\begin{split}\label{LIMITS1}
	E_1(\phi)  \to m a_0 \delta( \phi) \;,\quad\quad 
	a_1(\phi) \to m a_0 H(\phi) \;.
\end{split}
\ee
Our chosen parameterisation corresponds to holding fixed, in this limit, the product of the peak electric field $E_0 \sim \omega a_0$, and the effective temporal duration~$\sim1/\omega$. In other words we fix the total \textit{work done} by the field $m a_0 = eE_0/ \omega$; thus $a_0$ matches the usual definition of the dimensionless intensity parameter~\cite{RitusReview,DiPiazza:2011tq}. In order to evaluate the amplitude for pair production in this limit, it is helpful to make the boundary terms in~(\ref{DC-REG-1}) explicit. We thus rewrite (\ref{DC-REG-1}) as
\be\label{DC-REG-2}
	\mathcal{M} = - e^{i\Phi} \frac{\text{Spin}(a)}{i\Phi'}\bigg|^{X}_{-X} + \int\limits_{-X}^{X}\!\ud\phi \, e^{i\Phi} \text{Spin}(a) \;,
\ee
in which $X$ is an arbitrary positive constant. In the limit of interest, the Volkov solutions (see~\cite{volkov35,RitusReview,DiPiazza:2011tq,King:2015tba,Seipt:2017ckc}) yield the following explicit expressions for $\Phi$ and Spin, with $a_\mu = m a_0 \delta_\mu^1$ now constant,
\be
	\Phi = \begin{cases}
		\displaystyle 	\frac{l.q}{n.l-n.q} \phi \;, & \phi > 0 \\
		\displaystyle 	\frac{l.\bar\pi}{n.l-n.q}\phi\,, & \phi < 0
		\end{cases}
		\;, \qquad 
		\text{Spin}(a) =  \begin{cases}
			\displaystyle  \bar{u}_p \slashed{\epsilon} v_q & \phi > 0 \\
			\displaystyle  \bar{u}_p (1-\frac{ \slashed{a}\slashed{n}}{2 n.p} )\slashed{\epsilon} (1+\frac{ \slashed{n} \slashed{a}}{2 n.q}) v_q  & \phi < 0 \;,
		\end{cases}
\ee
in which the phase is expressed in terms of 
\be\label{pi-bar-med-dc}
	\bar{\pi}^\mu := q^\mu - a^\mu + n^\mu \frac{2q.a-a^2}{2n.q} \;.
\ee
The interpretation of ${\bar\pi}_\mu$ is that it is the \textit{initial} momentum of a positron which has traversed the delta function pulse and has final momentum $q_\mu$. With these results, the $\ud\phi$-integral in (\ref{DC-REG-2}) can be performed by splitting the integration into the ranges $\phi>0$ and $\phi<0$. The boundary terms at $\pm X$ are then cancelled, and we lose (as we should) all dependence on the arbitrary $X$. What remains is
\be\label{M-typ}
	\mathcal{M} \to  \frac{\text{Spin}_<}{i\Phi^\prime_<} -\frac{\text{Spin}_>}{i\Phi^\prime_>} \;,
\ee
in which the subscripts $\gtrless$ refer to evaluation at $\phi>0$ and $\phi<0$, where both Spin and $\Phi$ are constants. This compact and exact result is easily understood; it means that we only pick up contributions from \textit{across} the delta function at the origin, where the arguments of $\Phi$ and Spin jump. This is typical of situations with delta functions, and is physically sensible as it means we pick up contributions only from where the fieldstrength is nonzero.

From here we mod-square, and average (sum) over initial (final) polarisations and spins. The resulting expression for the pair production probability is, in terms of the positron's final momentum components $q_\LCperp$ and $u:= n.q/n.l$, 
\be\label{DC-PAR-1}
	\mathbb{P} = \frac{\alpha m^2}{4\pi^2} \int\! \ud^2 q_\LCperp\!  \int\limits_0^1\!\frac{\ud u}{u}  \, (1-u) \bigg[  \frac{1}{(l.\bar\pi)^2}+ \frac{1}{(l.q)^2}   - \frac{2}{ l.{\bar \pi}  l.q}\big(1+a_0^2h(u)\big)\bigg] \;,
\ee
in which $h(u) = 1/2 - 1/4 u(1-u)$. The result is, as may be expected, very similar to that obtained for scattering of a particle off an instantaneous kick, as used to exemplify infra-red divergences of QED~\cite{Peskin:1995ev}. (There is no infra-red divergence here.) Fig.~\ref{FIG:EXACT-DC}, left hand panel, shows an example of the differential emission probability
\be\label{dP}
	\frac{u}{1-u}\frac{\ud^3 \mathbb{P}}{\ud^2q_\LCperp \ud u} \;.
\ee
In the figure we take $u=1/2$, the `symmetric point' which corresponds to the positron and electron each carrying half of the initial photon's lightfront momentum $n.l$. Taking $l_\LCperp=0$, describing a head-on collision between the photon and laser, we observe two peaks in the poistron spectrum, at momenta   
$q_\LCperp = 0$ and $q_\LCperp = a_\LCperp$. A `semiclassical' explanation of this result is as follows. Assume that all particles are created at zero transverse momentum. Then positrons created in the rise of the ultra-short pulse (of which we are taking a limit) are accelerated by practically the whole field and so  pick up the full possible transverse momentum $a_\LCperp$ from it. Hence the spectral peak at $q_\LCperp = a_\LCperp$. Positrons created as the field falls, on the other hand, see little of it and therefore pick up little momentum after creation. This gives the spectral peak at $q_\LCperp = 0$.

\subsection{Total probability}
\begin{figure}[t!]
\includegraphics[width=0.45\textwidth]{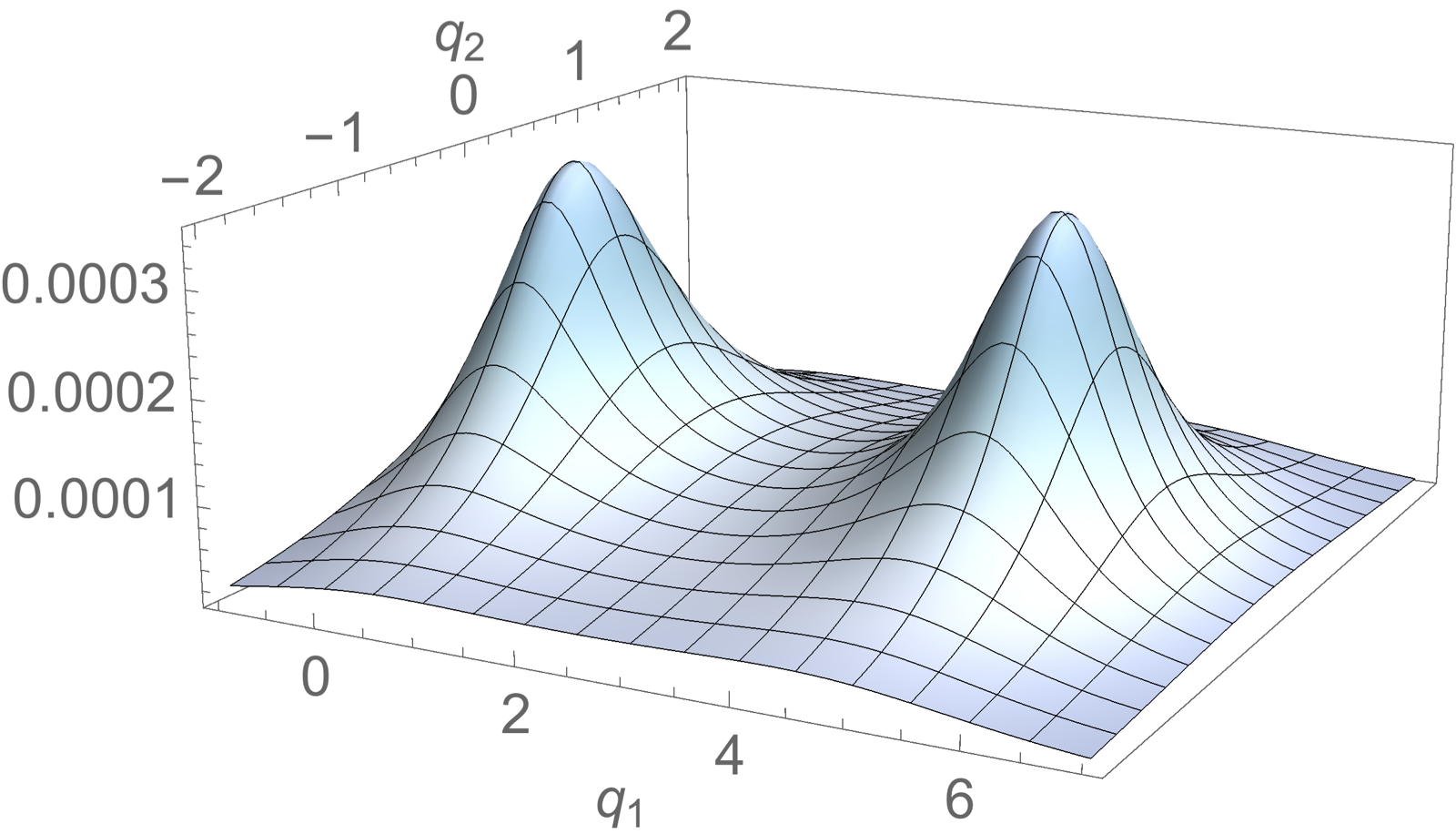} \hfill
\includegraphics[width=0.45\textwidth]{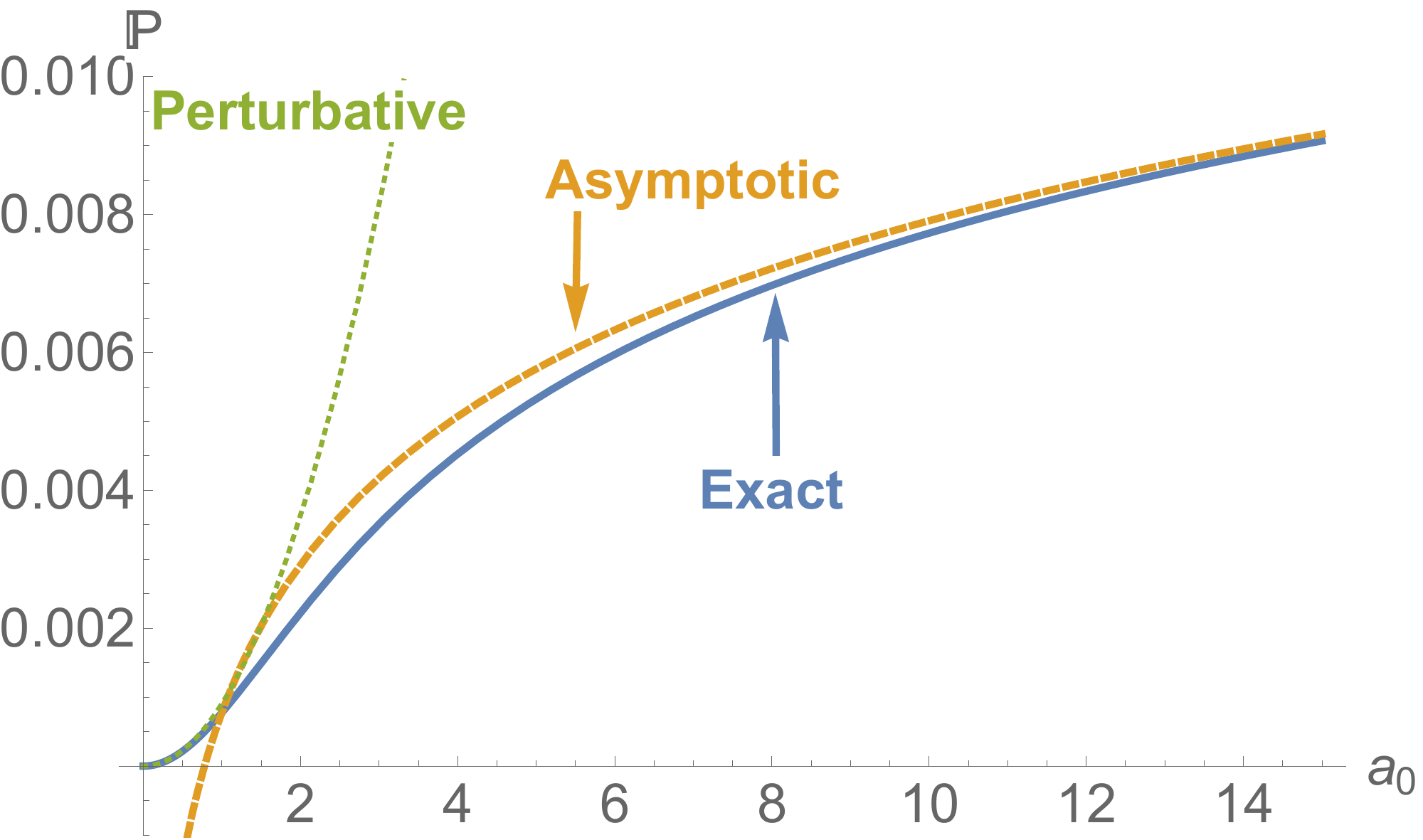}
\caption{\label{FIG:EXACT-DC} \textit{Left:} Differential emission probability (\ref{dP}) in the delta pulse, as a function of transverse momenta (all in units of electron mass $m$) at the symmetric point $u=1/2$, with $a_0=5$. The two peaks correspond to pairs produced (at zero transverse momentum) in the instantaneous rise and instantaneous fall of the field. \textit{Right}: the total pair creation probability, along with its small $a_0$ and large $a_0$ approximations.}
\end{figure}

In contrast to other cases, we can here also perform all the final state integrals in (\ref{DC-PAR-1}) and so obtain a closed form expression for the total pair production probability.  Consider the three terms in large square brackets in (\ref{DC-PAR-1}). In the first term, make a change of variables from $q_\LCperp$ to $v_\LCperp$ defined by $q_\LCperp = m v_\LCperp + u l_\LCperp + a_\LCperp$. Then that term becomes
\be
	\frac{\alpha}{\pi^2} \int\! \ud^2 v_\LCperp \int\limits_0^1\! \ud u\,  \frac{\, u(1-u)}{(1+|v_\LCperp|^2)^2} = \frac{\alpha}{6\pi} \;.
\ee
This is independent of $a_0$. A similar change of variable shows that the second term in (\ref{DC-PAR-1}) gives the same contribution. For the final term in (\ref{DC-PAR-1}), we make the same change of variable as here. The $u$-integral is then trivial, the angular integration in the $v_\LCperp$ plane may be performed using half-angle substitution, see~\cite[\S 3.613]{gradshteyn07}, and the $|v_\LCperp|$ integral can then be performed exactly (see below for details). The total probability is
\be\label{P-DC-EXAKT}
	\mathbb{P} = \frac{\alpha}{3\pi} + \frac{\alpha}{3\pi}\frac{ a_0^2-1}{a_0 \sqrt{a_0^2+4}} \log \bigg[1+\frac{a_0^4}{2}+2 a_0^2+a_0\sqrt{a_0^2+4} +\frac{1}{2} a_0^3\sqrt{a_0^2+4} \bigg] \;.
\ee
This exact result is plotted in Fig.~\ref{FIG:EXACT-DC}, right hand panel. We observe that the  probability is completely independent of the initial photon energy. This reflects the fact that, because the delta-pulse contains all (arbitrarily high) frequency modes which can contribute their energy to the process, a photon of any energy is capable of producing pairs in the background. While our result is exact for a delta-function pulse, it is also a first approximation to the probability in ultra-short pulses; we conclude that in such pulses the total probability of pair production will depend only weakly on initial photon energy.  (Our result does not imply a nonzero probability in the absence of the initial photon; we have assumed at almost all stages of the calculation that $l_\mu\not = 0$ (i.e.~when dividing by $l.q$, when introducing $u=n.q/n.l$, and so on). Expanding (\ref{P-DC-EXAKT}) for small $a_0$ shows that the probability behaves as
\be
	\mathbb{P} \simeq \frac{7}{18}\frac{\alpha}{\pi}a_0^2 \;, \quad a_0\ll 1 \;.
\ee
By the optical theorem, the above results extend to the imaginary part the photonic forward scattering amplitude at \textit{one loop}. This brings us to the question of how the probability (\ref{P-DC-EXAKT}) behaves at high intensity, or large $a_0$. Processes in constant crossed fields (the zero frequency limit of plane waves) can scale with powers of $a_0^{2/3}$~\cite{Ritus1,Morozov1,Narozhnyi:1979at,Narozhnyi:1980dc,Morozov2}, which has lead to a conjecture that for sufficiently high $a_0$ the Furry expansion as used here breaks down. If one invokes the ``locally constant field approximation'' (LCFA)~\cite{RitusReview} then the power-law scaling generalises to more general fields~\cite{Podszus:2018hnz,Ilderton:2019kqp}, but due to the many shortcomings of the LCFA~\cite{Harvey:2014qla,MeurenLCFA,LCFAPLUS,Podszus:2018hnz,Ilderton:2019kqp,King:2019cpj} it would be desirable to have exact analytic results. As we have a closed form result, (\ref{P-DC-EXAKT}), we may simply expand it for large $a_0$, finding
\be\label{hehheh}
		\mathbb{P} \simeq \frac{\alpha}{3\pi} + \frac{4}{3} \frac{\alpha}{\pi} \log a_0 \;, \quad a_0 \gg 1 \,.
\ee
Thus the probability scales logarithmically, asymptotically; there is no power-law scaling as in constant crossed fields (or as predicted by the LCFA, which is clearly not valid for short pulses). It is interesting that this exact result shows a logarithmic behaviour more typical of QED. The small and large $a_0$ approximations are plotted in Fig.~\ref{FIG:EXACT-DC}. Despite the complexity of the argument of the logarithm in (\ref{P-DC-EXAKT}), the exact result is very well approximated by the large $a_0$ expansion (\ref{hehheh}) for $a_0>10$.

\subsection{Quantum interference effects in pair production}\label{SECT:UTAN}
%
\begin{figure}[t]
\includegraphics[width=0.3\textwidth]{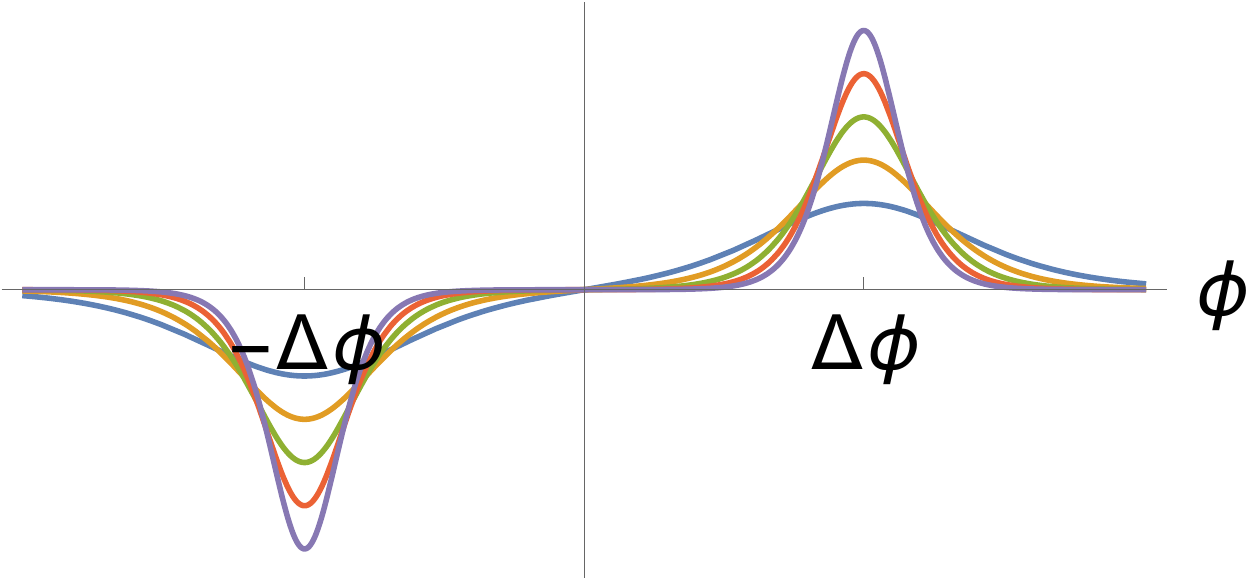} \qquad \qquad \qquad \qquad
\includegraphics[width=0.3\textwidth]{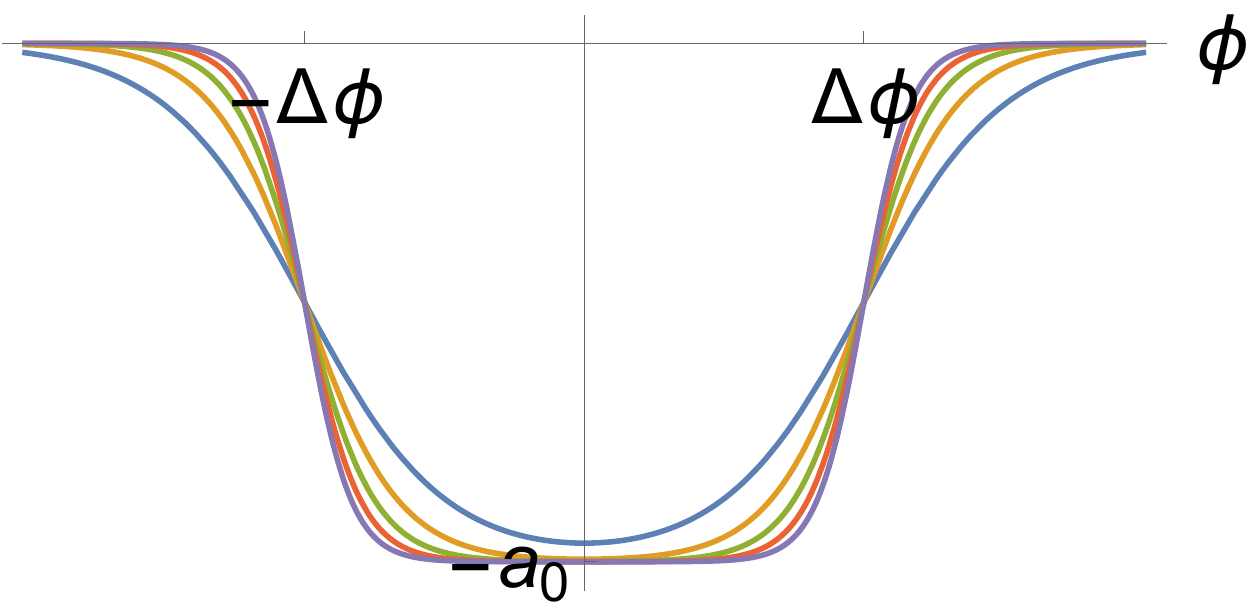}
\caption{\label{FIG:UTAN-DC} \textit{Left}: the electric field which, in the considered limit, becomes a sequence of two delta functions of opposite sign. \text{Right}: in the same limit, the potential becomes a box function with support on $|\phi|<\Delta\phi$.}
\end{figure}
\begin{figure}[t!]
\includegraphics[width=0.47\textwidth]{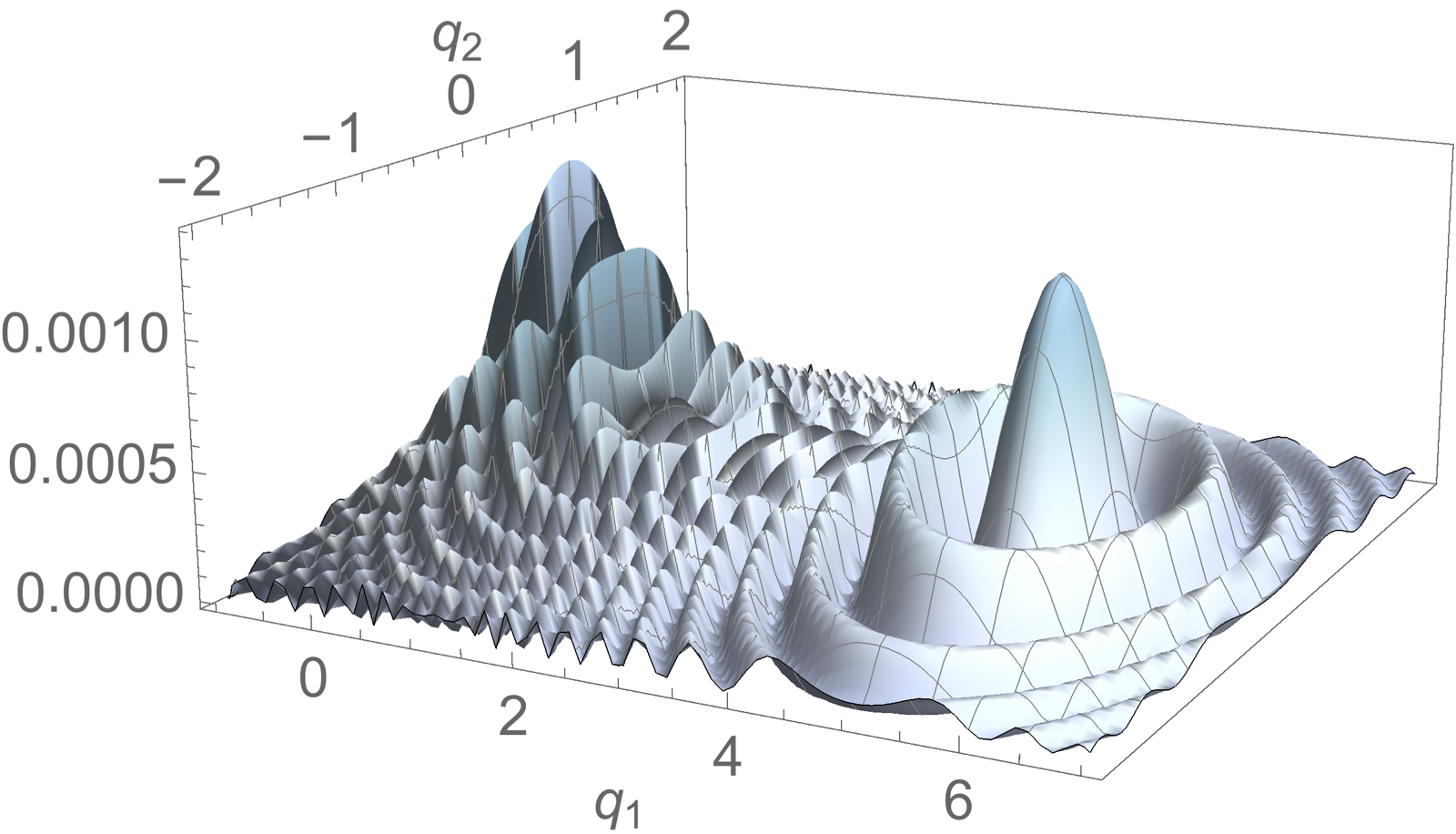}
\caption{\label{FIG:3D} Differential emission probability (\ref{dP}) as a function of transverse momenta (all in units of electron mass $m$) at $u=1/2$, with $a_0=5$, $\theta=1$, in two delta pulses, the first negative and the second positive. The spectrum is identically equal to that in a single positive delta peak, multiplied by the $4\sin^2$ interference term.}
\end{figure}

We turn now to a background composed of two delta pulses, as in (\ref{LIMITS1}), of opposite sign and separated by a (lightfront) time delay of $2\Delta\phi$, see Fig.~\ref{FIG:UTAN-DC}. In this case the field has no DC component, meaning that, classically (and ignoring radiation and back reaction), a particle traversing the whole field would pick up zero net energy-momentum. This is usually assumed when modelling laser pulses, which are naturally oscillating fields. In this case, the pair production amplitude is found to be
\be\begin{split}\label{17}
	\mathcal{M} &\to \bigg(\int\limits_{-\Delta\phi}^{\Delta\phi}\!\ud\phi \, e^{i\Phi}\bigg) \bigg[\text{Spin}(a) - \frac{l.\bar\pi}{l.p}\text{Spin}(0)\bigg]   = 2\sin\bigg[\frac{ \Delta \phi\,  l.\bar\pi}{n.l(1-u)}\bigg]\bigg(\frac{\text{Spin}_<}{\Phi'_<} - \frac{\text{Spin}_>}{i\Phi^\prime_>}\bigg) \;, 
\end{split}
\ee
in which all the variables, $\bar\pi$, $\Phi_\gtrless$ and so on, are exactly as for the single delta pulse above. It follows that the emission spectrum and probability are given simply by inserting the factor $4\sin^2$ under the integrals of (\ref{DC-PAR-1}): 
\be\label{P-PAR-UTAN-DC}
	\mathbb{P} = \frac{\alpha m^2}{4\pi^2} \int\! \ud^2 q_\LCperp \int\limits_0^1\!\frac{\ud u}{u}  \, (1-u)\, 4 \sin^2\bigg[\frac{\Delta \phi\, l.\bar\pi}{n.l(1-u)}\bigg] \bigg(  \frac{1}{(l.\bar\pi)^2} +\frac{1}{(l.q)^2}  - \frac{2}{l.q l.{\bar \pi} }\big[1+a_0^2h(u)\big]\bigg) \;.
\ee
The difference compared to the case of a single delta pulse is the $4\sin^2$ interference factor, which represents a coherent enhancement of the rate through quantum interference~\cite{Hebenstreit:2009km,Akkermans:2011yn} (to be discussed in detail elsewhere). The pair spectrum (\ref{dP}) is illustrated in Fig.~\ref{FIG:3D}.  Note that the spectral peaks remain at the same positions $q_\LCperp = 0$ and $q_\LCperp = a_\LCperp$ as before. Using the above semiclassical picture and referring to Fig.~\ref{FIG:UTAN-DC}, positrons created in the rise of the first peak are accelerated and decelerated by the whole field, picking up $-a_\LCperp$ in transverse momentum at the first delta, and then $+a_\LCperp$ at the second, thus ending with zero transverse momentum. Positrons created in the fall of the second peak receive no momentum from the field. These two creation events interfere and source the spectral peak at $q_\LCperp=0$. Positrons created in the fall of the first delta, or the rise of the second, pick up only $+a_\LCperp$ in transverse momentum from the second delta. These two events interfere and source the second spectral peak at $q_\LCperp= a_\LCperp$.

The degree of interference in the spectrum is controlled by both the separation of the delta pulses and the energy of the initial photon through the dimensionless parameter $\theta:= m^2\Delta \phi / n.l$. For small $\theta$ the pair production probability is suppressed, because (see Fig.~\ref{FIG:UTAN-DC}) the field vanishes in this limit. For $\theta$ large, the differential spectrum exhibits extremely rapid oscillations due to interference effects. In the same limit, these effects drop out of the \textit{total} probability as follows; make the same change of variables from $q_\LCperp$ to $v_\LCperp$ as above, then the interference factor becomes
\be\label{4till2}
	4 \sin^2\bigg[\frac{\theta}{2}\frac{1+v^2}{u(1-u)}\bigg] = 2 - 2  \cos\bigg[\theta\frac{1+v^2}{u(1-u)}\bigg]\;.
\ee
The argument of the cosine is at least $4\theta$ and (as will be made clear below) the remaining $v$-integral in $\mathbb{P}$ is dominated by contributions from $v\sim a_0$; hence if $a_0$ is large and $\theta$ not too small, then to a first approximation the cosine factor is very rapidly oscillating, and integrating over it will give zero contribution to the total probability. What remains is the leading $2$ in (\ref{4till2}); hence, in the large $a_0$ limit, interference effects drop out of the \textit{integrated} probability, which becomes approximately equal to twice that in a single delta ~(\ref{hehheh}). We will see in the next section that there are processes for which interference effects `persist', leading to a different high intensity scaling in the total probability.

\section{Nonlinear Compton scattering} \label{SECT:NLC}
We now consider nonlinear Compton scattering (NLC), that is the emission of a photon, momentum $l_\mu$, from an electron of momentum $p_\mu$ incident on an intense field. We will express the probability in terms of the produced photon's transverse momentum $l_\LCperp$ and its longitudinal momentum fraction $s := n.l/n.p$. We begin again with a single delta pulse as in Fig.~\ref{FIG:MED-DC}. This field is unipolar, which means the NLC emission probability will show the usual logarithmic divergence of QED in the infra-red~\cite{Dinu:2012tj}. Due to the symmetry properties of the plane wave, it is convenient to regulate this divergence using a small cutoff $\varepsilon_\text{IR}$ in the lightfront momentum fraction $s$ of the emitted photon, rather than in its frequency. With this, a similar calculation to that for pair production leads to the emission probability 
\be\label{NLC-MED-DC}
	\mathbb{P}_\text{\tiny NLC} = -\frac{\alpha m^2}{4\pi^2} \int\!{\ud^2 l_\LCperp} \int\limits_{\varepsilon_\text{IR}}^1 \!\frac{\ud s}{s} \, (1-s) \bigg[\frac{1}{(l.\pi)^2} + \frac{1}{(l.p)^2} - \frac{2}{l.\pi l.p}\big(1+a_0^2g(s)\big)\bigg] \;,
\ee
in which, for $a_\mu = ma_0 \delta_{\mu}^1$ again,
\be\label{pi-igen}
	\pi^\mu = p^\mu + a^\mu + \frac{-2a.p - a^2}{2n.p} n^\mu \;, \qquad g(s) := \frac{1}{2} + \frac{1}{4}\frac{s^2}{1-s} \;.
\ee
The calculation of the integrals in (\ref{NLC-MED-DC}) differs from those for (\ref{P-DC-EXAKT}) only in the $s$-integration, which is trivial. We therefore quote the exact final result, dropping only terms which vanish as $\varepsilon_\text{IR}\to 0$:
\be\label{NLC-UNI}
	\mathbb{P}_\text{\tiny NLC} = \frac{2\alpha}{\pi}(1+ \log \varepsilon_\text{\tiny IR}) -\frac{2\alpha}{\pi}\frac{  \log \big[\text{pair}\big]}{  a_0 \sqrt{a_0^2+4}}\big( 1 +\tfrac{3}{8} a_0^2 + (1 + \tfrac12 a_0^2) \log \varepsilon_\text{\tiny IR}\big) \;,
\ee
in which ``pair'' indicates precisely the same argument as for pair production (\ref{P-DC-EXAKT}). The essential difference is clearly only in the dependence on the small~$s$ cutoff. The leading behaviour for $a_0\gg 1$ is again logarithmic,
\be
	\mathbb{P}_\text{\tiny NLC} \sim \frac{2\alpha}{\pi}(1+\log \varepsilon_\text{\tiny IR})   - \frac{4\alpha}{\pi} (\tfrac34 + \log \varepsilon_\text{\tiny IR})\log a_0 \;.
\ee
By the optical theorem, and replacing $\varepsilon_\text{\tiny IR}$ with a detector resolution $\varepsilon_\text{min}$, the NLC probability contributes to the one-loop forward scattering amplitude through $\mathbb{P}_\text{foward} = 1-\mathbb{P}_\text{\tiny NLC}$.

\subsection{Total emission probability in alternating sign pulses}
As for pair production, we also consider two alternating sign pulses as in Fig.~\ref{FIG:UTAN-DC}. In this case there is no infra-red divergence. A direct calculation shows that the total probability is again given by inserting a $4\sin^2$ interference factor into the single-pulse result (\ref{NLC-MED-DC}):
\be\label{NLC-UTAN-DC}
	\mathbb{P} = -\frac{\alpha m^2}{4\pi^2} \int\!{\ud^2 l_\LCperp} \int\limits_0^1 \!\frac{\ud s}{s} \, (1-s) 4\sin^2\bigg[\frac1b\frac{ l.\pi}{m^2(1-s)}\bigg]\bigg( \frac{1}{(l.\pi)^2} +\frac{1}{(l.p)^2}   - \frac{2}{l.\pi l.p}\big[1+a_0^2g(s)\big]\bigg) \;,
\ee
in which $1/b = \Delta\phi m^2/n.p$ is the analogue of the interference-energy parameter in pair production. ($\pi^\mu$ as defined above may here be interpreted as the Lorentz force momentum of the incoming electron, momentum $p_\mu$, between the first and second deltas.) There are clear similarities with pair production, so we focus on the differences, one of which is the high-intensity scaling of the probability.  The key observation is the following. We cannot, as we did for pair production, write  $4\sin^2x \to 2 - 2\cos 2x$ and drop the second term as rapidly oscillating. The physical reason is that doing so would \textit{introduce} a divergence at $s=0$, because the replacement would, in effect, reduce the probability to that in a \textit{single} delta pulse.

To see what is happening, we change variables in (\ref{NLC-UTAN-DC}), defining $v_\LCperp := l_\LCperp/s - \pi_\LCperp$, and we express the integral in terms of the modulus $v$ and polar angle $\vartheta$ of $v_\LCperp$. We consider the final term in the large round brackets of (\ref{NLC-UTAN-DC}) which (we will see) is the dominant term. In terms of our new variables the integral to be performed is
\be
\label{DEL-TRE-0}
\begin{split}
	\mathbb{P} &= \frac{8\alpha}{\pi^2}\int\limits_0^\infty\!\frac{\ud v\, v}{1+v^2} \int\limits_0^1\! \frac{\ud s}{s}(1-s) 
	\big(1+a_0^2 g(s)\big) \sin ^2\bigg[\frac{s (1+v^2 )}{2 b (1-s)}\bigg] \int\limits_{0}^{2\pi}\!\ud\vartheta \frac{1}{1+v^2 +a_0^2-2a_0 v \cos \vartheta } + \ldots\;.
\end{split}
\ee
Consider the $\sin^2$ factor; independent of the value of $b$ and $v^2$, the factor $s/(1-s)$ can still be arbitrarily \textit{small}, and hence there is a portion of the integration range (small $s$) over which $\sin^2$ oscillates arbitrarily \textit{slowly}, in contrast to pair production. This demonstrates why we cannot simplify $4\sin^2$ to an overall $2$. Now, we again use half-angle substitution to perform the $\vartheta$-integral,
\be
\label{DEL-TRE}
\begin{split}
	\mathbb{P} &=  \frac{16\alpha}{\pi}\int\limits_0^\infty\!\frac{\ud v\,}{1+v^2} 
	\frac{v}{\sqrt{a_0^4-2 a_0^2(v^2-1)+(1+v^2)^2}}
	\int\limits_0^1\! \frac{\ud s}{s}(1-s) 
	\big(1+a_0^2 g(s)\big) \sin ^2\bigg[\frac{s (1+v^2 )}{2 b (1-s)}\bigg] + \ldots
\end{split}
\ee
The $s$-integral can be performed exactly, but is an unenlightening combination of Sine- and Cosine-integrals. However, this function is well approximated by its asymptotic expansion for large argument, i.e. when $x \gtrsim 1$,
\be\label{app-0}
	\int\limits_0^1\! \frac{\ud s}{s}(1-s) 
	\big(1+a_0^2 g(s)\big) \sin ^2\bigg[\frac{s\, x}{2(1-s)}\bigg] \simeq \frac{1}{4} a_0^2 \big(\log (x)+ \gamma_E -\tfrac34 \big)+\frac{1}{2} \big(\log (x)+\gamma_E -1\big) \;. 
\ee
It is sufficient to use this approximation in establishing the leading order behaviour for $a_0\gg1$, for the following reasons. The relevant argument for us is $x = (1+v^2)/b$ which is at least equal to $1/b$. Inspection of (\ref{DEL-TRE}) shows that the remaining $v$-integral is strongly peaked around $v\simeq1$ and $v\simeq a_0$. The height of the former peak is independent of $a_0$, whereas the latter increases with $a_0$ and hence gives the leading order large $a_0$ contribution; this behaviour is the same when using (\ref{app-0}). Hence we substitute (\ref{app-0}) into (\ref{DEL-TRE}). Define $a_* = \sqrt{a_0^2+4}$ (which has already appeared in (\ref{P-DC-EXAKT})) and change variables from $v$ to $t$ defined by $1+v^2=a_0 a_* / t$; with this, the integrals to be performed are
\be
\label{DEL-TRE-almost}
\begin{split}
	\mathbb{P} &\sim  \frac{8\alpha}{\pi a_0 a_*} \int\limits_0^{a_0 a_*}\! \frac{\ud t }{\sqrt{1+t^2-2t a_0/a_*}} 	\Big( k_1 - k_2 \log t\Big) 
=: \frac{8\alpha}{\pi a_0 a_*} \big( k_1 I_1 - k_2 I_2) \;,
\end{split}
\ee
which defines the integrals $I_1$ and $I_2$, and where the constants $k_j$ are determined by (\ref{app-0}) to be, writing down only the dominant contributions,
\be\label{kk}
	k_2 \simeq \tfrac14 a_0^2  \;, \qquad \qquad k_1  \simeq \tfrac{1}{4}a_0^2 \log \frac{a_0a_*}{b}  \sim \tfrac12 a_0^2 \log a_0 \quad \text{for} \quad a_0 \gg 1. 
\ee
The integral $I_1$ may be calculated exactly (and is also that needed for pair production in a single delta pulse):
\be
	I_1 = \log \Big[1+a_0 a_* + \frac{a_0^3 a_*}{2}+\frac{a_0^2 a_*^2}{2}\Big] \sim 4 \log a_0 \quad \text{for} \quad  a_0\gg 1. 
\ee
We have not found a closed form expression for the remaining integral $I_2$. However, we can find the leading order behaviour for large $a_0$, as follows. As $a_0\to\infty$, the ratio $a_0/a_*$ under the square root approaches unity at leading order, and the integrand jumps at $t=1$ but remains integrable. In this limit the only dependence on $a_0$ is in the upper integral limit. Hence we can obtain the leading order contribution by approximating
\be\begin{split}\label{Li}
	I_2  \to \int\limits_0^{a_0^2} \!\ud t\,  \frac{\log{t}}{|1-t|}  = -\frac{\pi^2}{6} - \text{Li}_2(1-a_0^2) \sim 2 \log^2a_0 \;,
\end{split}
\ee
in which Li is the polylogarithm. We are finally in a position to evaluate (\ref{DEL-TRE-almost}). Combining (\ref{kk})--(\ref{Li}) we obtain
\be\label{loglog}
	\mathbb{P} \sim  \frac{8\alpha}{\pi a_0^2}\bigg( \frac12 a_0^2 \log a_0 \cdot  4 \log a_0  - \tfrac14 a_0^2 \cdot 2 \log^2 a_0 \bigg) = \frac{12\alpha}{\pi} \log^2 a_0 + \ldots
	\ee
This result in fact gives the asymptotic scaling of the \textit{full} probability. The first term in (\ref{NLC-UTAN-DC}) is independent of $a_0$. The calculation of the second term proceeds similarly to that above; the $\vartheta$-integral is very similar, the $s$-integral is the same, and may again be approximated by its large argument expansion. The resulting $v$-integral can also be performed, and the resulting contribution goes like $\log(a_0^2/b)$, which (like all the energy dependence) is subleading compared to (\ref{loglog}). The difference in scaling compared to pair production (double rather than single logarithm) is due to the different infra-red behaviours of the two processes; in effect, it is not possible to completely suppress interference effects in NLC, and this is what changes the asymptotic scaling.

The obtained scaling is also quite remarkable in comparison to other results in the literature. For constant crossed fields, the asymptotic scaling is $\mathbb{P} \sim (a_0^2 /b)^{1/3}$, and the same is implied to hold for pulses by invoking the LCFA.  We have now seen explicitly, for both nonlinear Compton scattering and pair production, as well as the one-loop effects connected to them via the optical theorem, that there is no such scaling in the ultra-short pulses considered here. 

\section{Conclusions}\label{SECT:CONCS}
We have considered scattering processes in strong plane wave backgrounds. It is well known that the amplitudes for such processes contain highly oscillatory, cumbersome phase integrals which require numerical integration. By using delta functions to model the limit of ultra-short pulses, we have been able to perform these integrals exactly, and so evaluate without approximation the emission spectra of nonlinear Breit Wheeler pair production and nonlinear Compton scattering. We have also explained the main features of the emission spectrum in terms of an intuitive physical picture. We remark that the extension of our calculations to sequences of three or more delta pulses is straightforward; the $S$-matrix element will always be composed of sums of terms like (\ref{M-typ}), from jumps across the deltas, multiplied by accumulated phases, as in e.g.~(\ref{17}).

In some cases it is also possible to perform all final state momentum integrals exactly, and thus obtain exact, closed form expressions for total scattering probabilities as a function of initial parameters, in particular the field intensity. We have seen in particular that the power-law scaling with intensity inferred from constant crossed field results is completely absent, replaced with a logarithmic scaling.   For both pair production and nonlinear Compton, it was the study of the low (lightfront) energy part of the emission spectrum which allowed us to establish the high-intensity scalings; it is interesting to note that this is precisely the part of the spectrum where the locally constant field approximation fails~\cite{Harvey:2014qla}. The potential implications for the NR conjecture will be discussed elsewhere. 

While constant fields are unphysical in the infra-red, our delta pulses should be expected to be unphysical in the UV. (This is however not an issue for the high \textit{intensity} scaling in which we are primarily interested here~\cite{Podszus:2018hnz,Ilderton:2019kqp}.)  One goal for future work is therefore the identification of other backgrounds, neither infinitely long nor short, which also allow all integrals to be performed analytically~\cite{Seipt:2016rtk}.  Finding such examples may be challenging, given the complexity evident in e.g.~(\ref{P-DC-EXAKT}) even for the case of delta-function pulses with little internal structure. However, such examples would provide us with improved insight into physics in strong fields, allow us to better check numerical methods and approximations~\cite{Harvey:2014qla,MeurenLCFA,LCFAPLUS}, and could also provide particular data to help analyse the double copy~\cite{Bern:2008qj,Bern:2010ue,Bern:2010yg,Carrasco:2015iwa,BernRev} of gauge theory on background plane waves~\cite{Adamo:2017nia,Adamo:2018mpq,Adamo:2019zmk}.  It would also be very interesting to see how much progress can be made with challenging higher-order (many vertex) processes in delta-pulse backgrounds.

\acknowledgments

A.I.~is supported by The Leverhulme Trust, project grant RPG-2019-148.

\end{document}